\pdfoutput=1
\documentclass[         %
aps,                    
jpa,                    
showpacs,               
nofootinbib,            
showkeys,               %
preprintnumbers,        %
floatfix]               
{revtex4}               
\tighten
\usepackage{graphicx} 

\begin{document}

\title{Towards a very precise knowledge of $\theta_{13}$}


\author{A.B. Balantekin}
  
  \address{Physics Department, University of Wisconsin, Madison WI 53706 USA}

\begin{abstract}
Recent experimental developments towards obtaining a very precise value of the third neutrino mixing angle, $\theta_{13}$, are summarized. Various implications of the measured value of this angle are briefly discussed.  
\end{abstract}

\maketitle


\section{INTRODUCTION}

The distinction between electroweak eigenstates and the mass eigenstates of neutrinos, i.e. neutrino mixing, is experimentally well-established. First indications of non-zero values of these mixing angles came from the observation of the solar neutrino deficit. The angle $\theta_{12}$ was first measured with solar neutrino detectors and later more precisely with reactor antineutrino experiments. The angle $\theta_{23}$ was measured  from the oscillation of neutrinos produced in the upper atmosphere by cosmic rays. In contrast, the remaining mixing angle, $\theta_{13}$, was poorly known. For a while it was thought that $\theta_{13}$ may even be zero, since a null value implies a peculiar symmetry between second and third generations (see e.g. Ref. \cite{Balantekin:1999dx}). However, as various neutrino experiments moved from the discovery stage to the precision stage, the slight tension between solar and reactor neutrino experiments hinted a non-zero value of $\theta_{13}$ \cite{Balantekin:2008zm,Fogli:2008jx}. A few years later, indications of electron neutrino appearances from 
accelerator-produced off-axis muon neutrinos at the T2K experiment suggested a relatively large value of $\theta_{13}$ \cite{Abe:2011sj}. 
Soon after the first T2K results, Daya Bay reactor neutrino experiment provided the first direct measurement of this angle \cite{An:2012eh}, followed by the RENO reactor neutrino experiment \cite{Ahn:2012nd}. As the multi-detector configuration at the Daya Bay experiment gathers more data,  
very soon $\theta_{13}$ will be the best known neutrino mixing angle. A comparison of the results from three reactor experiments is shown in Figure 1 and the current value of $\sin^2 2 \theta_{13}$ measured at various experiments is shown in Table \ref{tab:a}.

\begin{figure}[b]
\caption{(Color online) Comparison of the three reactor experiments. Statistical errors are shown in heavier lines (black online) and the systematic ones are shown in lighter lines (red online).}
\includegraphics[height=.3\textheight]{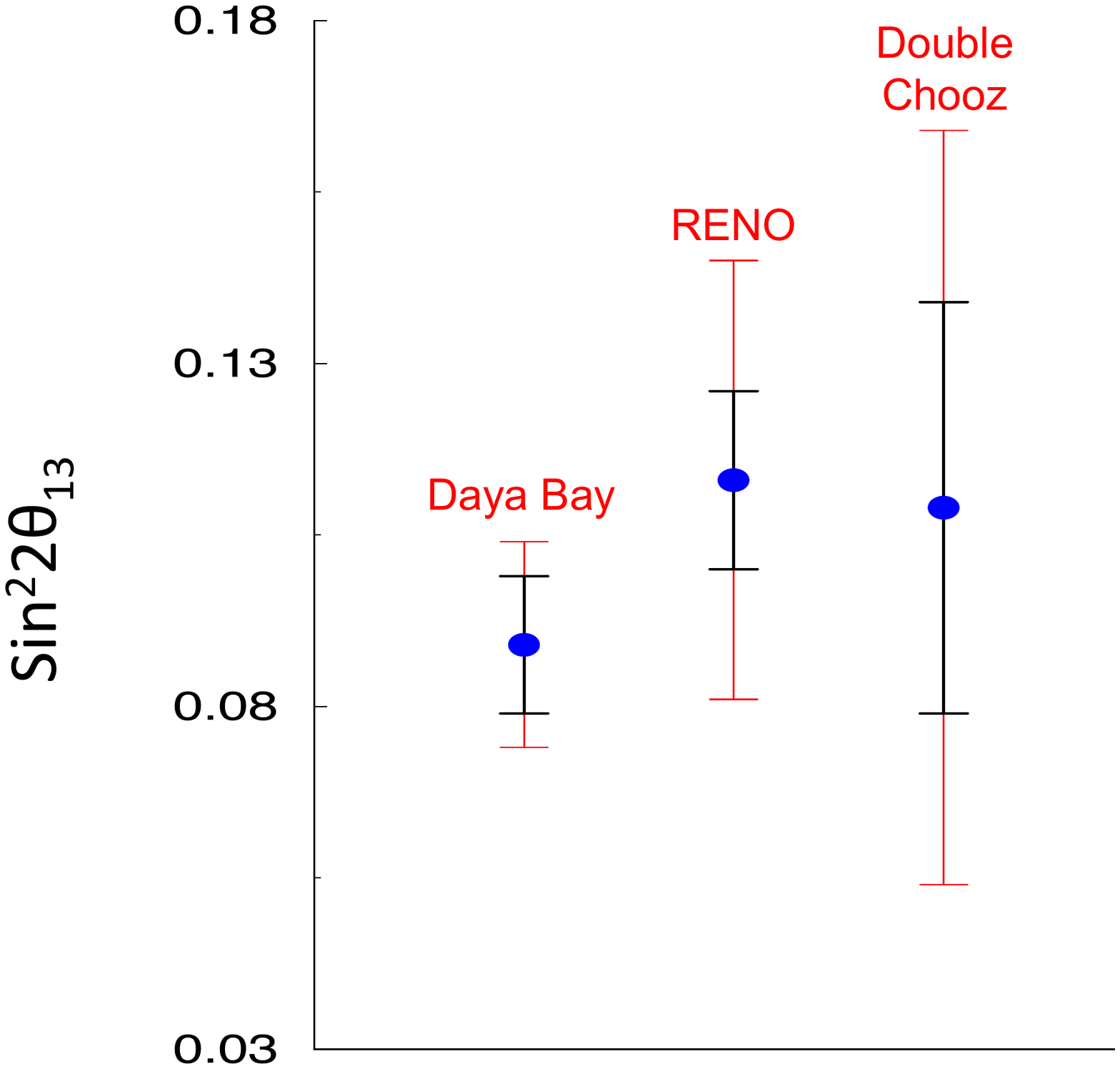}
\end{figure}

\begin{table}
\begin{tabular}{lr}
\hline
{Experiment}
  & {Measured ${\mathbf \sin^2 2 \theta_{13}}$} \\
\hline
Daya Bay \cite{:2012bu}  & $0.089\pm 0.010({\rm stat})\pm0.005({\rm syst})$\\
RENO \cite{Ahn:2012nd}  & $0.113 \pm 0.013 ({\rm stat}) \pm 0.019 ({\rm syst})$\\
Double Chooz \cite{Abe:2012tg} &    $0.109 \pm 0.030 ({\rm stat}) \pm 0.025 ({\rm syst})$       \\
T2K (assuming $\theta_{23} = \pi/4$) \cite{t2k} & $0.104 + 0.060 - 0.045 $\\
\hline
\end{tabular}
\caption{Values of $\sin^2 2 \theta_{13}$ measured at various experiments.}
\label{tab:a}
\end{table}

The neutrino mixing matrix connects mass eigenstates to flavor eigenstates:
\begin{equation}
\label{1}
| \nu_{\rm flavor} \rangle = {\bf T} |\nu_{\rm mass} \rangle. 
\end{equation}
We adopt the following parametrization of this matrix: 
\begin{equation}
\label{1a}
{\bf T} =  {\bf T}_{23}{\bf T}_{13}{\bf T}_{12}  = 
\left(
\begin{array}{ccc}
 1 & 0  & 0  \\
  0 & C_{23}   & S_{23}  \\
 0 & -S_{23}  & C_{23}  
\end{array}
\right)
\left(
\begin{array}{ccc}
 C_{13} & 0  & S_{13} e^{-i\delta_{CP}}  \\
 0 & 1  & 0  \\
 - S_{13} e^{i \delta_{CP}} & 0  & C_{13}  
\end{array}
\right) 
\left(
\begin{array}{ccc}
 C_{12} & S_{12}  & 0  \\
 - S_{12} & C_{12}  & 0  \\
0  & 0  & 1  
\end{array}
\right)
\end{equation}
where $C_{ij} = \cos \theta_{ij}$, $S_{ij} = \sin \theta_{ij}$, and $\delta_{CP}$ is the CP-violating phase.  Using the evolution operator, ${\bf U}$, which satisfies the evolution equation
\begin{equation}
\label{2}
i \frac{\partial}{\partial t} {\mathbf U} = {\mathbf H} {\mathbf U}, 
\end{equation}
the probability amplitude for transitions between flavor eigenstates $\alpha$ and $\beta$ can be written as 
\begin{equation}
\label{3}
A (\nu_{\alpha} \rightarrow \nu_{\beta} ) = {\rm Tr} ( {\mathbf U} {\mathbf R}^{(\alpha \beta)}),  
\end{equation}
where {\bf R} is the appropriate projection operator. 
For the electron neutrino survival probability this projection operator is given as 
\begin{equation}
\label{4}
{\mathbf R}^{(ee)} = \left( 
\begin{array}{ccc}
 1 & 0  & 0  \\
 0 & 0  & 0  \\
  0&0   & 0  
\end{array}
\right) , 
\end{equation}
and for the $\nu_{\mu} \rightarrow \nu_e$ transition as 
\begin{equation}
\label{5}
{\mathbf R}^{(\mu e)} = \left( 
\begin{array}{ccc}
 0 & 0  & 0  \\
 1 & 0  & 0  \\
  0&0   & 0  
\end{array}
\right) . 
\end{equation}
For vacuum oscillations one has
\begin{equation}
\label{6}
{\mathbf U} = {\mathbf T} 
\left(
\begin{array}{ccc} 
 \exp (-i E_1t) & 0  & 0  \\
 0& \exp (-iE_2t) & 0  \\
  0&0   & \exp (-iE_3t)
\end{array}
\right) 
{\mathbf T}^{\dagger} ,
\end{equation}
where $E_i = \sqrt{ p^2 + m_i^2}$. Noting
\begin{equation}
\label{7}
{\mathbf R}^{(ee)} {\mathbf T}_{23} = {\mathbf R}^{(ee)}
\end{equation}
one sees that the mixing angle $\theta_{23}$ drops out of the electron neutrino survival probability\footnote{This is still true if the matter effects are also included \cite{Balantekin:2003dc}.}. The electron neutrino survival probability then takes the form
\begin{equation}
\label{8}
P (\nu_e \rightarrow \nu_e) = 1 - \sin^2 2 \theta_{13} \left[ \cos^2 \theta_{12} \sin^2( \Delta_{31} L) + \sin^2 \theta_{12} \sin^2 (\Delta_{32} L) 
\right] - \cos^4 \theta_{13} \sin^2 2 \theta_{12} \sin^2 (\Delta_{21} L) 
\end{equation}
where we defined
\begin{equation}
\label{9}
\Delta_{ij} = \frac{\delta m_{ij}^2}{4E} = \frac{m_i^2-m_j^2}{4E}. 
\end{equation}
Note that the sign of $\Delta_{ij}$ is controlled by the neutrino mass hierarchy. 

\section{Appearance versus disappearance experiments}

Reactor experiments aiming to measure $\theta_{13}$ detect the loss of the original electron antineutrino flux: they are disappearance experiments integrating over muon- and tau-neutrino channels. The disappearance probability can be easily written from Eq. (\ref{8}) as 
\begin{equation}
\label{10}
1- P (\nu_e \rightarrow \nu_e) = \sin^2 2 \theta_{13} \left[ \cos^2 \theta_{12} \sin^2( \Delta_{31} L) + \sin^2 \theta_{12} \sin^2 (\Delta_{32} L) 
\right] + \cos^4 \theta_{13} \sin^2 2 \theta_{12} \sin^2 (\Delta_{21} L) .
\end{equation}
Clearly if one is at close enough distances to the reactor, $\Delta_{21} L \sim 0$ and the "solar" oscillation represented by the second term can be ignored. Using the fact $\Delta_{32}  \sim \Delta_{31} $, suggested by the solar neutrino experiments, one obtains the disappearance probability
\begin{equation}
\label{11}
1- P (\nu_e \rightarrow \nu_e) = \sin^2 2 \theta_{13} \sin^2( \Delta_{31} L) . 
\end{equation}
Hence the very short-baseline reactor neutrino experiments, such as Daya Bay, Double Chooz, and RENO, unambiguously measure $\theta_{13}$ without needing a knowledge of other neutrino mixing angles or the mass hierarchy. It should be emphasized that multiple detector configurations currently employed in Daya Bay and RENO experiments also obviate any need for the knowledge of original reactor neutrino flux. 

The situation is rather different for experiments with longer baselines 
looking at the appearance of electron neutrinos in a flux of muon neutrinos. 
One difference is that since neutrinos travel trough Earth, matter effects need to be included. Another difference comes from the presence of other mixing angles. The resulting expressions are typically very complicated, but can be calculated in a series expansion 
\cite{Akhmedov:2004ny}. 
The appropriate appearance probability in the lowest order is 
\begin{equation}
\label{12}
P (\nu_{\mu} \rightarrow \nu_e) \sim \frac{ \sin^2 2 \theta_{13} \sin^2 \theta_{23}}{(1- G_F N_e /\sqrt{2} \Delta_{31})^2}  
\sin^2 \left[ \left ( \Delta_{31} - \frac{G_F N_e}{\sqrt{2}} \right) L \right] + {\cal O} (g) ,
\end{equation}
where we defined 
\begin{equation}
\label{13}
g = \frac{\delta m_{21}^2}{\delta m_{31}^2} \sim 0.03 .
\end{equation}
The next order correction in $g$ to the electron neutrino appearance probability 
brings in an additional dependence on the CP-violating phase in the neutrino mixing matrix. The denominator of the term multiplying the oscillating term in Eq. (\ref{12}) depends on the sign of $\delta m^2$, i.e. the mass hierarchy of the neutrinos. 
Consequently, appearance experiments such as T2K \cite{t2k} and MINOS \cite{Adamson:2011qu} cannot disentangle $\theta_{13}$ from other observables such as the mass hierarchy, the other mixing angle $\theta_{23}$ and the CP-violating phase. 
Note that, at least in principle, a medium baseline ($L \sim 60$ km) reactor antineutrino experiment will also have some sensitivity to the neutrino mass hierarchy 
\cite{Batygov:2008ku}. 

A detailed study investigating the physics potential of the experiments with a broad set of different beam, near- and far-detector configurations  is presented in Ref. \cite{Akiri:2011dv}.  

\section{Implications of the measured value of $\theta_{13}$}

\subsection{Solar neutrinos}

To extract $\theta_{13}$ from the solar neutrino data many times the formula
\begin{equation}
\label{b1}
P_{3\times3}( \nu_e \rightarrow  \nu_e) = \cos^4{\theta_{13}} \> 
P_{2\times2}( \nu_e \rightarrow  \nu_e \>{\rm with}\> N_e
\cos^2{\theta_{13}})  + \sin^4{\theta_{13}} 
\end{equation}
is used. This formula is obtained by expanding the full survival probability as a power series in $\sin \theta_{13}$. (For different derivations see 
Refs. \cite{Balantekin:2011ta} and \cite{Fogli:2001wi}). Now that we know the actual value of $\theta_{13}$, we could ascertain how good an approximation results from Eq. (\ref{b1}).
 
Corrections to this formula can be calculated as  \cite{Balantekin:2011ta,Fogli:2001wi}
\begin{equation}
\label{a18}
P_{3\times3}( \nu_e \rightarrow  \nu_e) = \cos^4{\theta_{13}}  ( 1 - 4 \sin^2\theta_{13} \alpha) \> 
P_{2\times2}( \nu_e \rightarrow  \nu_e \>{\rm with}\> N_e
\cos^2{\theta_{13}})  + \sin^4{\theta_{13}}  (1+ 4 \cos^2 \theta_{13} \alpha). 
\end{equation} 
where we introduced 
\begin{equation}
\label{14}
\alpha = \frac{1}{\Delta_{32} + \Delta_{31}} \sqrt{2} G_F N_e (r=0) . 
\end{equation}
The value of the quantity $\alpha$ increases with increasing neutrino energy. Hence to calculate the largest correction we calculate it at 10 MeV, where the flux of the highest energy $^8$B solar neutrinos is maximal.  
For a 10 MeV neutrino with $\delta m_{32}^2 \sim 2 \times 10^{-3}$ eV$^2$, assuming an electron density of $\sim$100 N$_{\rm A}$/cm$^3$ in the neutrino production region, the expansion parameter $\alpha$ is rather small: 
\begin{equation}
\label{15}
\alpha \sim 3 \times 10^{-2}. 
\end{equation}
Using the recently measured value of $\theta_{13}$ at Daya Bay, we can calculate the corrections to the order $\alpha$ in Eq. (\ref{a18}). For the term multiplying $\cos^4{\theta_{13}}$ we get a correction of 0.27\% and for the  term multiplying $\sin ^4{\theta_{13}}$ we can a correction of about 11.7\%. However this correction changes the latter term from $5.2 \times 10^{-4}$ to $5.8 \times 10^{-4}$ and can be safely ignored. We conclude that calculations of the solar neutrino properties in the literature using Eq. (\ref{b1}) can be considered as 
fairly reliable. 

\subsection{Supernova neutrinos}

The location of the MSW resonance depends on the neutrino mass-squared differences. This puts the resonance governed by $\delta m_{21}^2$ at solar densities. For the resonance governed by  $\delta m_{31}^2$, the appropriate matter density is a bit higher and matter with this higher density exists in the outer shells of a supernova. 
Core-collapse supernovae are likely sites for several nucleosynthesis scenarios. One of these is nucleosynthesis via neutrino-induced nucleon emission (the $\nu$-process). 
Rather general considerations, independent of the detailed dynamics of supernovae, suggest a hierarchy of energies of neutrinos emitted from the proto-neutron star, namely 
$E_{\nu_e} < E_{\bar{\nu}_e} < E_{\nu_{\mu},\nu_{\tau},{\bar{\nu}}_{\mu},{\bar{\nu}}_{\tau}}$ \cite{Balantekin:2003ip}. 
In the presence of non-zero values of $\theta_{13}$, the average energy of the electron antineutrinos taking part in the charged-current antineutrino reactions is 
smaller for a normal mass hierarchy than for an inverted hierarchy. For the normal hierarchy electron antineutrinos undergo no MSW resonance and energy hierarchy mentioned above holds. However for the inverted hierarchy, if $\theta_{13}$ is large enough, MSW resonance is operational for antineutrinos in the outer shells of the supernova and the hotter muon and tau antineutrinos can be converted into electron antineutrinos,
Consequently it was pointed some time ago that the synthesis of $^{11}$B and 
$^7$Li via in $\nu$-process in supernovae is sensitive to the neutrino mass hierarchy for not too small values of $\theta_{13}$ 
\cite{Yoshida:2005uy}: $^{11}$B is mainly produced through the neutral current reaction sequence   
$^4$He($\nu,\nu'$p)$^3$H($\alpha,\gamma$)$^7$Li($\alpha,\gamma$)$^{11}$B and the charged-current reaction sequence 
$^4$He($\bar{\nu}_e,e^+$n)$^3$H($\alpha,\gamma$)$^7$Li($\alpha,\gamma$)$^{11}$B. A small amount (12 to 16\%) of the $^{11}$B can also be produced in the He layer of the supernova from $^{12}$C through the neutral-current reaction $^{12}$C($\nu,\nu'$p)$^{11}$B and the charged-current reaction $^{12}$C($\bar{\nu}_e,e^+$p)$^{11}$B. Because of the presence of the MSW effect for the inverted hierarchy, muon and tau antineutrinos can be converted into more energetic electron antineutrinos, boosting the 
nucleosynthesis yields outlined above. Clearly the yield of the $^{11}$B, synthesized in the $\nu$-process, would then depend on the value of $\theta_{13}$. Of course, the detection of the $^7$Li and $^{11}$B abundances in the supernova material enriched by the $\nu$-process is itself a very difficult task. However, it  had been suggested some time ago that materials synthesized in a supernova may be trapped within the SiC grains in carbonaceous chondrite meteorites \cite{nuprocess}. It turns out that such chondrites are studied extensively by geologists \cite{mur}. A recent study 
\cite{fujiya} found that a particular meteorite may contain supernova-produced $^{11}$B and $^7$Li trapped in its grains. Motivated by this finding and the recent precise measurements of the $\theta_{13}$, the authors of Ref. \cite{Mathews:2011jq} 
carried out a Bayesian analysis of the uncertainties in the measured meteoritic material as well as uncertainties of the supernova model calculations. They found 
a marginal preference for the inverted hierarchy. 

In a core-collapse supernova environment neutrino-neutrino interactions are no longer negligible as the gravitational binding energy of the progenitor massive star is converted into $\sim 10^{58}$ neutrinos during the cooling process of the proto-neutron star\footnote{For recent reviews see Refs. \cite{Duan:2009cd}, \cite{Duan:2010bg}, and \cite{Raffelt:2010zza}}. 
The resulting collective neutrino oscillations play a crucial role both for neutrinos and antineutrinos. 
Since such collective neutrino oscillations dominate the neutrino propagation much deeper than the conventional matter-induced MSW effect, they would impact r-process nucleosynthesis yields if core-collapse supernovae are shown to be the appropriate sites.  The quantity that governs the 
yields from the r-process nucleosynthesis is the neutron-to-proton ratio (or equivalently the electron fraction). A preliminary investigation of the dependence of this ratio on the neutrino mixing angle $\theta_{13}$ was given in Ref. \cite{Balantekin:2004ug} 
using the mean field approximation to the collective neutrino oscillations. Although other rather interesting results already appeared in the literature \cite{Duan:2010af}, much work still needs to be done in this direction. 

Collective neutrino oscillations can also cause an interesting effect which may impact Li/B ratio, discussed above. One of the 
consequences of the collective neutrino oscillations is that, at a particular energy, final neutrino energy spectra are almost completely divided into parts of different flavors \cite{Raffelt:2007cb,Duan:2008za,Pehlivan:2011hp,Galais:2011gh}. 
These phenomena are called spectral splits (or swappings). Once the neutrinos reach the He shells of the supernovae, complete swappings between electron neutrinos (or antineutrinos) and other flavors would not be distinguishable from the adiabatic MSW oscillations \cite{Fogli:2007bk}. This suggests a re-analysis of the Li/B ratio taking into account both the matter-enhanced oscillations and the collective effects. 

\section{Conclusions}

The neutrino mixing angle $\theta_{13}$ is well on its way to be the best known mixing angle. Contrary to the naive expectations, it turned out to be relatively large (but not large enough to invalidate the perturbative expansion in powers of its sine, widely used in solar neutrino physics). 
The full implications of the measured value of this angle in supernova physics are yet to be completely explored. However 
this value, coupled with our understanding of supernova nucleosynthesis and the fossil record of the $\nu$-process produced $^7$Li and 
$^{11}$B, provides tantalizing hints about the neutrino mass hierarchy. 


\section*{acknowledgments}
This work was supported in part 
by the U.S. National Science Foundation Grant No.  PHY-1205024.
and
in part by the University of Wisconsin Research Committee with funds
granted by the Wisconsin Alumni Research Foundation. 
I would like to thank T. Kajino and G. Matthews for useful conversations. 
I also thank the Center for Theoretical Underground Physics and Related Areas (CETUP* 2012) in South Dakota for its hospitality and for partial support during the completion of this work. 


\bibliographystyle{aipprocl} 
\bibliography{sample}

\begin{thebibliography}{99}

\bibitem{Balantekin:1999dx} 
  A.~B.~Balantekin and G.~M.~Fuller,
  \emph{Phys.\ Lett.\ B} \textbf{471}, 195 (1999)
  [hep-ph/9908465].

\bibitem{Balantekin:2008zm} 
  A.~B.~Balantekin and D.~Yilmaz,
  \emph{J.\ Phys.\ G} \textbf{35}, 075007 (2008)
  [arXiv:0804.3345 [hep-ph]]; 

\bibitem{Fogli:2008jx} 
  G.~L.~Fogli, E.~Lisi, A.~Marrone, A.~Palazzo and A.~M.~Rotunno,
  \emph{Phys.\ Rev.\ Lett.}  \textbf{101}, 141801 (2008)
  [arXiv:0806.2649 [hep-ph]].

\bibitem{Abe:2011sj} 
  K.~Abe {\it et al.}  [T2K Collaboration],
  \emph{Phys.\ Rev.\ Lett.}  \textbf{107}, 041801 (2011)
  [arXiv:1106.2822 [hep-ex]].

\bibitem{An:2012eh} 
F.~P.~An {\it et al.}  [DAYA-BAY Collaboration],
  \emph{Phys.\ Rev.\ Lett.}  \textbf{108}, 171803 (2012)
  [arXiv:1203.1669 [hep-ex]]; 
 F.~P. An, {\it et al.}  [Daya Bay Collaboration],
  \emph{Nucl.\ Instrum.\ Meth.\ A} \textbf{685}, 78 (2012)
  [arXiv:1202.6181 [physics.ins-det]].

\bibitem{Ahn:2012nd} 
  J.~K.~Ahn {\it et al.}  [RENO Collaboration],
  \emph{Phys.\ Rev.\ Lett.}  \textbf{108}, 191802 (2012)
  [arXiv:1204.0626 [hep-ex]].

\bibitem{:2012bu} 
F.~P. An, {\it et al.}  [Daya Bay Collaboration],
  arXiv:1210.6327 [hep-ex]; 

\bibitem{Abe:2012tg} 
  Y.~Abe {\it et al.}  [Double Chooz Collaboration],
  \emph{Phys.\ Rev.\ D} \textbf{86}, 052008 (2012)
  [arXiv:1207.6632 [hep-ex]].
  
 \bibitem{t2k}
T.~Nakaya, for the TRK Collaboration, reported in Neutrino 2012. 

\bibitem{Balantekin:2003dc} 
  For a more detailed explanation see 
  A.~B.~Balantekin and H.~Yuksel,
  {\it J.\ Phys.\ G} \textbf{29}, 665 (2003)
  [hep-ph/0301072].

\bibitem{Akhmedov:2004ny} 
  E.~K.~Akhmedov, R.~Johansson, M.~Lindner, T.~Ohlsson and T.~Schwetz,
  {\it JHEP} {\bf 0404}, 078 (2004)
  [hep-ph/0402175].

\bibitem{Adamson:2011qu} 
  P.~Adamson {\it et al.}  [MINOS Collaboration],
  \emph{Phys.\ Rev.\ Lett.}  {\bf 107}, 181802 (2011)
  [arXiv:1108.0015 [hep-ex]].

\bibitem{Batygov:2008ku} 
  M.~Batygov, S.~Dye, J.~Learned, S.~Matsuno, S.~Pakvasa and G.~Varner,
  arXiv:0810.2580 [hep-ph]; 
  P.~Ghoshal and S.~T.~Petcov,
  {\it JHEP} {\bf 1103}, 058 (2011)
  [arXiv:1011.1646 [hep-ph]]; 
  S.~-F.~Ge, K.~Hagiwara, N.~Okamura and Y.~Takaesu,
  arXiv:1210.8141 [hep-ph].

\bibitem{Akiri:2011dv} 
  T.~Akiri {\it et al.}  [LBNE Collaboration],
  arXiv:1110.6249 [hep-ex].

\bibitem{Balantekin:2011ta} 
  A.~B.~Balantekin,
  {\emph J.\ Phys.\ Conf.\ Ser.}  {\bf 337}, 012049 (2012)
  [arXiv:1106.5021 [hep-ph]].

\bibitem{Fogli:2001wi}
  G.L.~Fogli, E.~Lisi and A.~Palazzo, 
  {\it Phys.\ Rev.\  D} {\bf 65}, 073019 (2002)
  [arXiv:hep-ph/0105080].

\bibitem{Balantekin:2003ip} 
   See e.g. 
  A.~B.~Balantekin and G.~M.~Fuller,
  J.\ Phys.\ G {\bf 29}, 2513 (2003)
  [astro-ph/0309519].

\bibitem{Yoshida:2005uy} 
  T.~Yoshida, T.~Kajino and D.~H.~Hartmann,
  \emph{Phys.\ Rev.\ Lett.}  {\bf 94}, 231101 (2005)
  [astro-ph/0505043].

\bibitem{nuprocess}
S. Amari, P. Hoppe, E. Zinner, and R.S. Lewis, \emph{Astrophys. J. Lett.} \textbf{394}, L43 (1992); 
L.R. Nittler, S. Amari, E. Zinner, S.E. Woosley, and R.S. Lewis, {\it ibid.} \textbf{462}, L31 (1996); P. Hoppe, {\it et al.}, 
\emph{Meteoritics and Planetary Sci.} \textbf{35}, 1157 (2000). 

\bibitem{mur}
A. Besmehn and P. Hoppe, \emph{Geochim. Cosmochim. Acta} \textbf{67}, 4693 (2003). 

\bibitem{fujiya}
W. Fujiya, P. Hoppe and U. Ott, \emph{Astrophys. J. Lett.} \textbf{730}, L7 (2011). 

\bibitem{Mathews:2011jq} 
  G.~J.~Mathews, T.~Kajino, W.~Aoki, W.~Fujiya and J.~B.~Pitts,
  \emph{Phys.\ Rev.\ D} {\bf 85}, 105023 (2012)
  [arXiv:1108.0725 [astro-ph.HE]].

\bibitem{Duan:2009cd} 
  H.~Duan and J.~PKneller,
  \emph{J.\ Phys.\ G} {\bf 36}, 113201 (2009)
  [arXiv:0904.0974 [astro-ph.HE]].

\bibitem{Duan:2010bg} 
  H.~Duan, G.~M.~Fuller and Y.~-Z.~Qian,
  \emph{Ann.\ Rev.\ Nucl.\ Part.\ Sci.}  {\bf 60}, 569 (2010)
  [arXiv:1001.2799 [hep-ph]].

\bibitem{Raffelt:2010zza} 
  G.~G.~Raffelt,
  \emph{Prog.\ Part.\ Nucl.\ Phys.}  {\bf 64}, 393 (2010).

\bibitem{Balantekin:2004ug} 
  A.~B.~Balantekin and H.~Yuksel,
  \emph{New J.\ Phys.}  {\bf 7}, 51 (2005)
  [astro-ph/0411159].

\bibitem{Duan:2010af} 
  H.~Duan, A.~Friedland, G.~C.~McLaughlin and R.~Surman,
  \emph{J.\ Phys.\ G} {\bf 38}, 035201 (2011)
  [arXiv:1012.0532 [astro-ph.SR]].

\bibitem{Raffelt:2007cb} 
  G.~G.~Raffelt and A.~Y.~.Smirnov,
  \emph{Phys.\ Rev.\ D} {\bf 76}, 081301 (2007)
  [Erratum-ibid.\ D {\bf 77}, 029903 (2008)]
  [arXiv:0705.1830 [hep-ph]].

\bibitem{Duan:2008za} 
  H.~Duan, G.~M.~Fuller and Y.~-Z.~Qian,
  \emph{Phys.\ Rev.\ D} {\bf 77}, 085016 (2008)
  [arXiv:0801.1363 [hep-ph]].

\bibitem{Pehlivan:2011hp} 
  Y.~Pehlivan, A.~B.~Balantekin, T.~Kajino and T.~Yoshida,
  \emph{Phys.\ Rev.\ D} {\bf 84}, 065008 (2011)
  [arXiv:1105.1182 [astro-ph.CO]].

\bibitem{Galais:2011gh} 
  S.~Galais and C.~Volpe,
  Phys.\ Rev.\ D {\bf 84}, 085005 (2011)
  [arXiv:1103.5302 [astro-ph.SR]].

\bibitem{Fogli:2007bk} 
  G.~L.~Fogli, E.~Lisi, A.~Marrone and A.~Mirizzi,
  JCAP {\bf 0712}, 010 (2007)
  [arXiv:0707.1998 [hep-ph]].

\end{thebibliography}

\end{document}